\documentclass[manuscript]{acmart}

\usepackage{tikz} 
\usetikzlibrary{positioning}
\usetikzlibrary{patterns}
\usepackage{caption}
\usepackage{subcaption}
\usepackage{amsthm}
\usepackage[linesnumbered,lined,ruled,noend]{algorithm2e}
\usetikzlibrary{positioning}

\AtBeginDocument{%
  \providecommand\BibTeX{{%
    \normalfont B\kern-0.5em{\scshape i\kern-0.25em b}\kern-0.8em\TeX}}}

\setcopyright{rightsretained}
\copyrightyear{2022}
\acmYear{2022}
\acmConference[RecSys '22]{Sixteenth ACM Conference on Recommender Systems}{September 18--23, 2022}{Seattle, USA}
\acmBooktitle{Sixteenth ACM Conference on Recommender Systems (RecSys '22), September 18--23, 2022, Seattle, USA}
\acmPrice{xx.xx}
\acmDOI{xx.xxxx/xxxxxxx.xxxxxxx}
\acmISBN{xxx-x-xxxx-xxxx-x/xx/xx}

\copyrightyear{2022} 
\acmYear{2022} 
\setcopyright{rightsretained}\acmConference[RecSys '22]{Sixteenth ACM Conference on Recommender Systems}{September 18--23, 2022}{Seattle, WA, USA}
\acmBooktitle{Sixteenth ACM Conference on Recommender Systems (RecSys '22), September 18--23, 2022, Seattle, WA, USA}
\acmPrice{xx.xx}
\acmDOI{xx.xxxx/xxxxxxx.xxxxxxx}
\acmISBN{xxx-x-xxxx-xxxx-x/xx/xx}

\begin{document}

\title{A User-Centered Investigation of Personal Music Tours}

\author{Giovanni Gabbolini}
\affiliation{
\institution{Insight Centre for Data Analytics, School of Computer Science \& IT}
\country{University College Cork, Ireland}
}
\email{giovanni.gabbolini@insight-centre.org}
\orcid{0000-0001-7914-9999}

\author{Derek Bridge}
\orcid{0000-0002-8720-3876}

\affiliation{\mbox{ }
\institution{Insight Centre for Data Analytics, School of Computer Science \& IT}
\country{University College Cork, Ireland}
}
\email{d.bridge@cs.ucc.ie}

\begin{abstract}
  Streaming services use recommender systems to surface the right music to users. Playlists are a popular way to present music in a list-like fashion, \ie{} as a plain list of songs. An alternative are \textit{tours}, where the songs alternate with \textit{segues}, which explain the connections between consecutive songs. Tours address the user need of seeking background information about songs, and are found to be superior to playlists, given the right user context. In this work, we provide, for the first time, a user-centered evaluation of two tour-generation algorithms (\greedy{} and \optimal{}) using semi-structured interviews. We assess the algorithms, we discuss attributes of the tours that the algorithms produce, we identify which attributes are desirable and which are not, and we enumerate several possible improvements to the algorithms, along with practical suggestions on how to implement the improvements.
  Our main findings are that \greedy{} generates more likeable tours than \optimal{}, and that three important attributes of tours are segue diversity, song arrangement and song familiarity. More generally, we provide insights into how to present music to users, which could inform the design of user-centered recommender systems.
\end{abstract}


\newcommand{\g}[1]{\textcolor{blue}{\textbf{[Giovanni] #1}}}
\newcommand{\db}[1]{\textcolor{red}{\textbf{[Derek] #1}}}
\newcommand{\q}[1]{\textbf{RQ#1}}
\newcommand{\p}[1]{$i_{#1}$}
\newcommand{\algoname}[1]{\textsc{{#1}}}
\newcommand{\dave}{\algoname{Dave}}
\newcommand{\greedy}{\algoname{Greedy}}
\newcommand{\hillclimbing}{\algoname{Hill-Climbing}}
\newcommand{\optimal}{\algoname{Optimal}}
\newcommand{\ie}{\textit{i.e.}}
\newcommand{\eg}{\textit{e.g.}}
\newcommand{\quot}[1]{``\textit{{#1}''}}
\renewcommand{\dots}[1]{[...]}
\newcommand{\rarity}{$\mathit{rarity}$}
\newcommand{\unpop}{$\mathit{unpopularity}$}
\newcommand{\shortness}{$\mathit{shortness}$}
\newcommand{\interestingness}{$\mathit{interestingness}$}
\newtheorem{problem}[theorem]{Problem}

\newcommand{\VAR}[1]{}
\newcommand{\BLOCK}[1]{}


\maketitle

\section{Introduction} \label{sec:intro}

\begin{figure*}
     \centering
     \begin{subfigure}[b]{0.3\textwidth}
         \centering
         \begin{tikzpicture}[
    tracknode/.style={rectangle, thick, align=center, draw=blue!60},
        node distance=0.3cm
    ]
    
    \node[tracknode] (song0) {\textit{Interstellar Love} by \textit{Thundercat}};
    
    \node[tracknode] (song1) [below=of song0] {\textit{Post Requisite} by \textit{Flying Lotus}};
    
    \node[tracknode] (song2) [below=of song1] {\textit{Wisdom Eye} by \textit{Alice Coltrane}};
    
    \draw[->, thick] (song0.south) -- (song1.north);
    \draw[->, thick] (song1.south) -- (song2.north);
    
    \end{tikzpicture}
         \caption{}
         
     \end{subfigure}
     \hfill
     \begin{subfigure}[b]{0.3\textwidth}
         \centering
         \begin{tikzpicture}[
        seguenode/.style={rectangle, thick, align=center, pattern=north west lines, pattern color=orange!30, draw=orange!60},
        tracknode/.style={rectangle, thick, align=center, draw=blue!60},
        node distance=0.3cm
        ]
        
        \node[tracknode] (song0) {\textit{Interstellar Love} by \textit{Thundercat}};
        
        \node[seguenode] (segue0) [below=of song0] {``Interstellar Love was \\ produced by Flying Lotus''};
        \node[tracknode] (song1) [below=of segue0] {\textit{Post Requisite} by \textit{Flying Lotus}};
        
        \node[seguenode] (segue1) [below=of song1] {``Flying Lotus is the grand-nephew \\ of Alice Coltrane''};
        \node[tracknode] (song2) [below=of segue1] {\textit{Wisdom Eye} by \textit{Alice Coltrane}};
        
        \draw[->, thick] (song0.south) -- (segue0.north);
        \draw[->, thick] (segue0.south) -- (song1.north);
        
        \draw[->, thick] (song1.south) -- (segue1.north);
        \draw[->, thick] (segue1.south) -- (song2.north);
        
    \end{tikzpicture}
         \caption{}

     \end{subfigure}
     \hfill
     \begin{subfigure}[b]{0.3\textwidth}
         \centering
         \begin{tikzpicture}[
        seguenode/.style={rectangle, thick, align=center, pattern=north west lines, pattern color=orange!30, draw=orange!60},
        tracknode/.style={rectangle, thick, align=center, draw=blue!60},
        node distance=0.3cm
        ]
        \node[tracknode] (song0) {\textit{Post Requisite} by \textit{Flying Lotus}};
        \node[seguenode] (segue0) [below=of song0] {``Flying Lotus is the grand-nephew \\ of Alice Contrane''};
        
        \node[tracknode] (song1) [below=of segue0] {\textit{Wisdom Eye} by \textit{Alice Coltrane}};
        \node[seguenode] (segue1) [below=of song1] {``Alice Coltrane has done  Jazz, \\ \& Thundercat has done Acid Jazz, \\ a genre that derives from Jazz''};
        
        \node[tracknode] (song2) [below=of segue1] {\textit{Interstellar Love} by \textit{Thundercat}};
        
        \draw[->, thick] (song0.south) -- (segue0.north);
        \draw[->, thick] (segue0.south) -- (song1.north);
        
        \draw[->, thick] (song1.south) -- (segue1.north);
        \draw[->, thick] (segue1.south) -- (song2.north);
        
    \end{tikzpicture}
         \caption{}
         
     \end{subfigure}
        \caption{A playlist of three songs (a), and two tours of three songs, recommended by \optimal{} (b) and \greedy{} (c). Shaded boxes contain segues; clear boxes contain songs.}
        \label{fig:toy_sample}
\end{figure*}
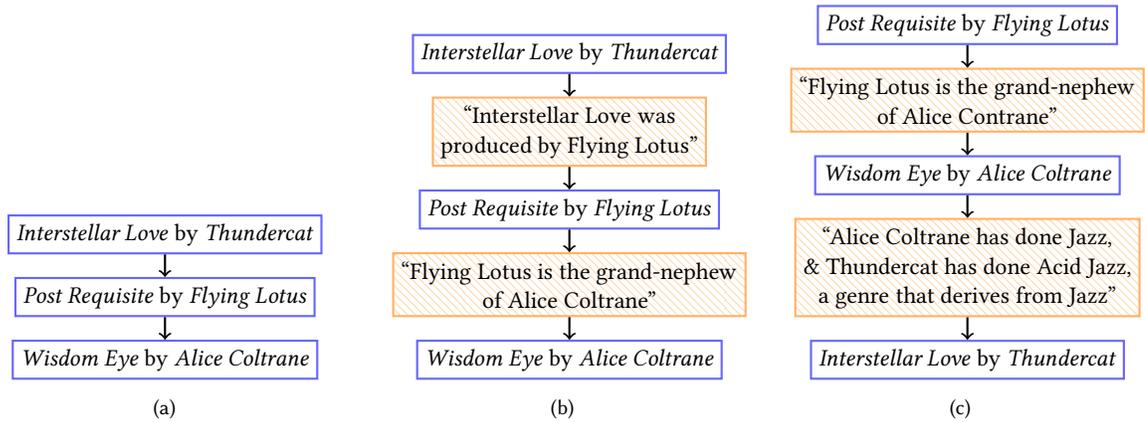

Playlists are a popular feature of music streaming services. Users consume playlists for 31\% of their total listening time \cite{schedl2018current}. Recommender systems are often used to surface the right playlists to users, \eg{} see \cite{bonnin2014automated} for a review. Playlists present music in a list-like fashion, \ie{} as a plain list of songs, such as in Figure \ref{fig:toy_sample} (a). Past research investigates how to present playlists best. For example, \cite{nakano2016playlistplayer, Pohle2007} propose to arrange songs according to audio similarity, such as beats per minute, and web-based similarity, such as  Wikipedia categories.

Playlists are limited in that they allow for little interaction with the user: one song follows the other, from the first to the last. Such a concept is suited for some occasions, such as passive listening, but is limiting in other contexts. For example, \cite{lee2011similar} finds that unfamiliar music in playlists should be explicitly connected to user background knowledge, so as to keep user interest alive. And, \cite{lonsdale2011we} finds that users seek background information while listening to music in order to stay informed and to build relationships with others.

An alternative to playlists are \textit{tours}, where the songs alternate with \textit{segues}, \ie{} short textual connections between songs, which explain the connection between consecutive songs, as in Figures \ref{fig:toy_sample} (b) and (c). Tours resemble the popular format of radio shows, where a presenter speaks and takes the audience from one song to another. As such, tours augment music with background information, and help overcome the above limitations of playlists. For example, \cite{Behrooz19} finds that tours offer a better user experience than playlists, given the right listening context.

Tour recommendation can be framed as a multi-objective problem \cite{rodriguez2012multiple}, as algorithms should surface the right songs, the right segues, and the right combination of the two. Existing algorithms tackle a simpler problem, in which the right songs are assumed to be known, and the task is to recommend segues, and an arrangement of the songs. 
We are aware of only two works on tours: \cite{Behrooz19} and \cite{gabbolini2021play}. The authors of \cite{Behrooz19} simplify the problem even further, as they assume they have the right arrangement (ordering) of the songs. They recommend segues only, by finding the shortest path in a graph where the nodes are songs, the edges are segues, and the edges' weights are assigned according to segue importance (defined manually by the authors) and segue diversity. The authors of \cite{gabbolini2021play} make a step forward, as they recommend both a song arrangement and the segues, so as to maximise the \textit{interestingness} of the segues in the tour. 

The two existing works on tours suffer from a number of limitations. In \cite{Behrooz19}, the authors set up semi-structured interviews to evaluate their algorithm, and the concept of tours in general. However, their study is limited in that they do not investigate three important topics for implementing a functioning tour recommender: (1) what is a good segue arrangement; (2) what is a good song arrangement; (3) what are the guidelines for distinguishing good from bad segues; and (4) what topics should segues cover.
The authors of \cite{gabbolini2021play} overcome limitation (3) by employing an interestingness measure, according to which segues can be scored for how good they are. The interestingness measure they use is validated in \cite{gabbolini2021generating}, where it is found to positively correlate with human perceptions of segue quality. Also, the authors of \cite{gabbolini2021play} partially overcome limitation (2), as they look to arrange songs so that the average interestingness of the segues in the tour is as high as possible. However, they do not discuss whether their arrangement strategy is a good one, especially from the user perspective. Finally, (1) and (4) are left totally unsolved by \cite{gabbolini2021play}. In fact, \cite{gabbolini2021play} does not make any consideration about what is a good segue arrangement, and they have segues covering different topics, but they do not validate these topics. Finally, both \cite{Behrooz19} and \cite{gabbolini2021play} are limited in that the songs themselves are not chosen by a recommender system, but assumed to be given, which would not be the case for a functioning tour recommender.

Here, we consider two algorithms proposed in \cite{gabbolini2021play}, \greedy{} and \optimal{}, and we set up semi-structured interviews, where interviewees judge tours recommended by the two algorithms. We omit \cite{gabbolini2021play}'s third algorithm, \hillclimbing{}, because \cite{gabbolini2021play} found it to produce tours mostly equivalent to those produced by \optimal{}, especially for small inputs. 
We help to overcome the limitations of current work on tours by investigating: what is a good segue arrangement; what is a good song arrangement; what topics should segues cover; and how to recommend the right music in tours. \greedy{} and \optimal{} are also evaluated in \cite{gabbolini2021play} via off-line experiments, and are found to differ in their total segue interestingness, and in the distribution of interestingness throughout the tour. With the interviews, we contribute to the first user-centered evaluation of the two algorithms. We are interested in checking the results of the off-line experiments, in assessing the overall quality of the tours recommended by the two algorithms, and the interestingness of their segues, all from the user perspective. Finally, we are interested in assessing whether users value the concept of tours in general, a fundamental issue already addressed by \cite{Behrooz19}, that we investigate further.

Summing up, we aim to fulfil five goals. \textbf{Goal (1)} is finding attributes of the tours. Attributes of interest could be: perceived interestingness distribution within tours; segue diversity; segue narrativity, defined as the quality that sequences of segues in a tour present a narrative with a coherent text; and song arrangement. \textbf{Goal (2)} is identifying what attributes of tours are desirable and not desirable. \textbf{Goal (3)} is assessing the quality of tours, in terms of segue interestingness and overall tour quality. \textbf{Goal (4)} is assessing whether users value the concept of tours in general. \textbf{Goal (5)} is identifying possible improvements to tours. 
In summary, we formulate the following research questions:

\begin{description}
\item[\q{1}:] What are the attributes of the tours?
\item[\q{2}:] What attributes of tours are good/bad?
\item[\q{3}:] Which algorithm recommends better tours, and why?
\item[\q{4}:] How valuable is the concept of tours in general? 
\item[\q{5}:] What are possible improvements to tours?
\end{description}



Our work adds 
actionable knowledge, 
which can inform functional tour recommender systems. More generally, we provide insights into how to present music to users, which could inform the design of user-centered music recommenders.


\section{Related work}
Music recommendation is a popular research topic with important real-world applications, such as in streaming services, where recommender systems are used to surface the right music to users. The music recommendation problem has unique characteristics compared to other content domains, such as books or movies \cite{schedl2015music}. For example, it is common to recommend lists of songs to users, or playlists, while it is less common to recommend lists of movies or books. As such, the playlist recommendation problem is a well-studied one. Bonnin and Jannach, for example, provide a review of algorithms for playlist recommendation \cite{bonnin2014automated}. Other research investigates more user-centered aspects. For example, \cite{nakano2016playlistplayer, Pohle2007} propose to arrange songs according to audio similarity, such as beats per minute, and web-based similarity, such as Wikipedia categories, while \cite{lee2011similar} studies user perception of the similarity among the songs in playlists, and gives insights on the ideal level of similarity, so as to inform the design of playlist recommenders. 

A related task to playlist recommendation is (automatic) playlist continuation (APC), which consists of automatically adding one or more tracks to a playlist in a way that fits the same target characteristics of the original playlist \cite{schedl2018current}. In the ``RecSys Challenge 2018'', participants were asked to implement an APC system able to continue real user playlists from the Spotify music streaming service \cite{chen2018recsys}. The challenge represents a landmark in APC research, and sets the state of the art to that date. The most competitive algorithms made use of simple Collaborative Filtering (CF) recommenders \cite{zamani2019analysis}, such as Matrix Factorisation \cite{koren2009matrix} and \textit{k}-Nearest Neighbors \cite{sarwar2001item} algorithms. More recently, deep learning APC algorithms were also proposed, \eg{} see \cite{tran2019adversarial}.
Our work studies an alternative to playlists: tours. In tours, the songs alternate with segues, which explain the connection between consecutive songs.

Segues were first introduced in \cite{Behrooz19}, and defined in \cite{gabbolini2021generating} as ``short texts that connect two items''. Both \cite{Behrooz19} and \cite{gabbolini2021generating} focus on 
the case where the items connected by the segues are songs. In \cite{Behrooz19}, the authors propose a tour recommendation algorithm where segues are chosen for their diversity, and importance, defined manually by the authors. In \cite{gabbolini2021generating}, the authors overcome the manual scoring, as they propose an interestingness measure, according to which segues can be scored for how interesting they are. Interestingness is based on statistical information, \eg{} the more rare a segue, the more interesting, and \eg{} the shorter the segue, the more interesting. The authors validate their interestingness concept in a user trial, and find that it is positively correlated with human perception of segue quality. In follow-up work \cite{gabbolini2021play}, the same authors utilise the interestingness measure to formulate three tour recommendation algorithms, which they evaluate via offline experiments. Our work in this paper builds on \cite{gabbolini2021play}, as we provide the first user-centered evaluation of the algorithms that they propose. Also, we investigate the user perspective of tours, and we provide practical and actionable knowledge which can inform the development of functional tour recommender systems.

\section{Method} \label{sec:method}
We employ two tour recommendation algorithms proposed in \cite{gabbolini2021play}, \greedy{} and \optimal{}, without any alteration. The algorithms assume that the set of songs to be included in the tour are given \textit{a priori}, and the algorithms must recommend segues and an arrangement of songs, so as to generate a tour of the kind shown in Figures \ref{fig:toy_sample} (b) and (c). 

Given a `small' set of songs, \eg{} several order of magnitudes smaller than the whole catalogue of songs, the goal of the algorithms is to find a permutation of the songs, and a list of segues between the songs, 
maximising the average $interestingness$ of the segues, \ie{} the $utility$.
The algorithms assume the existence of two functions: $segues$ and $interestingness$. We represent songs and information about songs in a knowledge graph; the $segues$ function finds paths in the graph between the two songs of interest, and  translates the paths to texts. As $interestingness$, we use the function proposed originally in \cite{gabbolini2021generating}, which is based on the infrequency and conciseness of knowledge graph paths before they are translated to segues. The graph contains a variety of information, such as: musical genres; locations, \eg{} where an artist was born; dates, \eg{} when an artist was born; record labels, \eg{} who published a song;  and relationships between artists, between songs, and between artists and songs. 
The knowledge graph is built from data harvested from two resources: Musicbrainz\footnote{\url{https://musicbrainz.org}} and Wikidata\footnote{\url{https://www.wikidata.org}}. In total, the knowledge graph features 30 distinct nodes types, and 205 distinct edge types. As such, segues cover a variety of different topics. Examples of segues recommended by the algorithms are in the the shaded boxes in Figures \ref{fig:toy_sample} (b) and (c). We provide a complete description of the nodes and edges that build up the knowledge graph in the additional materials\footnote{\url{https://doi.org/10.5281/zenodo.6817643}} and the implementation of the algorithms are freely available\footnote{\url{https://github.com/GiovanniGabbolini/play-it-again-sam}}.

\greedy{} is a heuristic algorithm, while \optimal{} always recommends the tour with highest $utility$. In particular, \greedy{} builds a solution iteratively, by choosing the next song to be the one with the segue of highest $interestingness$, while \optimal{} reduces the recommendation problem to a Traveling Salesman Problem, or TSP, and solves it optimally.

\section{Experiment}
\label{sec:experiment}
We set-up a semi-structured interview experimental protocol, as it suits our need to answer both open and closed questions in the same sitting \cite{adams2015conducting}. For example, one closed question is which of the tours produced by the two algorithms of Section \ref{sec:method} is preferred (part of \q{3}), while one open question is what can be improved in tours (\q{5}).

The interview protocol follows a within-subject design, \ie{} each participant is subject to both treatments (\greedy{} and \optimal{}), in order to elicit an explicit comparison of the algorithms. \greedy{} recommend tours which are different from those recommended by \optimal{}. For example, the first segue in tours by \greedy{} tends to be more interesting than the last, while in tours by \optimal{} the first and last segues tend to be equally interesting \cite{gabbolini2021play}.

The conversation revolves around example tours recommended by the algorithms.
Some days ahead of the interview, we ask participants to send us a list of ten songs they are familiar with, making sure they are all from different artists. We set the number of songs to ten so as to reduce fatigue effects.
We run the algorithms on the ten songs, and we prepare a \textit{pdf} textual file with a visual representation of the resulting tours, side by side, similar to Figure \ref{fig:toy_sample} (b) and (c). We randomise the order, so that sometimes \optimal{} is on the left hand-side and \greedy{} is on the right hand-side, and vice versa. We also prepare an \textit{mp3} audio file of the tour, comprising 15-second song previews, and segues that are read by a Text-to-Speech (TTS) engine, for a total duration of approximately seven minutes. Songs previews are from the Spotify API \footnote{\url{https://developer.spotify.com/documentation/web-api/libraries/}}, and the TTS engine is the Google Text-to-Speech engine\footnote{\url{https://pypi.org/project/gTTS/}}. We use songs previews rather than entire songs because previews are freely available. The first tour in the \textit{mp3} corresponds to the one that the participant will see on the left in the \textit{pdf}, and the second tour will be the one on the right in the \textit{pdf}. Both the \textit{mp3} and the \textit{pdf} files are used during the interview to support the conversation. Algorithm names are excluded from both the \textit{pdf} and \textit{mp3} files. A sample \textit{pdf} file and \textit{mp3} file used during an interview are in the additional materials.

This interview study is approved by our organisation's ethics committee and it is conducted one-to-one by an author (the interviewer) and a participant (the interviewee) via Zoom.

\subsection{Interviewees}
We recruit interviewees via a student mailing list within our university in Ireland. We make sure that the interviewee’s first language is English, so that they can properly understand the segues.
We ended up with 16 interviewees, who we will refer to as \p{1} to \p{16}. One interviewee is a graduate, while the other 15 are currently students. Among them, four are undergraduates and 11 are postgraduate students. The interviewees have/seek degrees in a range of disciplines: arts, psychology, medicine, geography, computer science, history, music, social sciences, food sciences and human-computer interaction. Also, 15 out of 16 interviewees are users of music streaming services.
We do not collect other demographic information, as we considered it not essential for the scope of the study.

\subsection{Interview guide} \label{sec:interview_guide}
At the beginning of the interview, the interviewee listens to the \textit{mp3} version of their two tours, and views the \textit{pdf}. Both the interviewer and the interviewee can hear the \textit{mp3} and both can see the \textit{pdf}. The \textit{pdf} remains visible to both, until the end of the interview. 
Once the \textit{mp3} is over, we ask questions aimed at answering the \q{}s of Section \ref{sec:intro}.

To answer \q{1-2}, we ask the interviewee to identify different attributes of the tours. We start with two open questions about what the interviewee likes and does not like about the tours, and we continue with other open questions probing for specific attributes of tours, such as distribution of segue interestingness throughout the tour (``is the first segue more interesting than the last?'', and ``does the interestingness of segues, from first to last, increase, decrease, stay equal, or something in the middle?''), segue diversity, segue narrativity (``are segues linked together to form a narrative, or are they independent pieces of text?''), segue top-bottom bias (``is the first segue more interesting than the last?'') and song arrangement.
Then, we ask the interviewer to assess each tour with respect to these attributes, and eventually express whether that attribute is important for the tours.
To answer \q{3}, we ask interviewees to assess the quality of the tours according to two dimensions: the interestingness of the segues and the overall tour quality. In particular, we ask interviewees to compare tours recommended by \greedy{} and tours recommended by \optimal{}, according to the two dimensions.
To answer \q{4}, it is necessary to take extra-care, as the interviewee might assume that tours are the work of the interviewer, and may tend to provide a positive concept evaluation accordingly. One strategy to relieve the pressure from the interviewee is showing non-judgemental acceptance, as suggested in \cite{adams2015conducting}. Therefore, to answer \q{4}, the interviewer asks ``People have mixed opinions on whether they would welcome a tour of their music or not. How do you see this issue?''. To answer \q{5}, the interviewer starts with two open questions about what the interviewee would like and wouldn't like to see in tours, and continues with open questions probing whether tours should feature non-familiar music or familiar music. We provide the full interview guide text in the additional materials.

\section{Results} \label{sec:results}
Our interviews produce a wealth of material: 11 hours of videos \ie{} 42 minutes per interviewee, on average. Interviews are manually transcribed by one of the authors, and we end up with a corpus of 44-thousand transcribed words.

We conduct a thematic analysis of our data to identify important ideas and themes from our interviews. Thematic analysis is used for many kinds of qualitative analysis work in human-computer interaction (HCI), \eg{} \cite{weisz2021perfection}. We follow the process described by Braun and Clarke \cite{braun2006using} in which researchers familiarize themselves with the data, then generate codes and group them to identify higher-level themes. Each phase of the analysis is conducted by one author independently, and validated among the authors.
We identify four main themes, \ie{} ``tour attributes'', ``tour quality'', ``concept evaluation'' and ``tour improvements''. The first theme addresses \q{1-2}, the second theme addresses \q{3}, the third theme addresses \q{4}, and the fourth theme addresses \q{5}. Every theme is covered in one Section below. Each theme has a number of sub-themes, covered in the  subsections.

\subsection{Tour attributes} \label{theme:1}

We ask interviewees to identify different attributes of the tours, to assess the tours with respect to each attribute, and eventually express whether that attribute is important for the tours, as we explained in Section \ref{sec:interview_guide}.
Each subsection below accounts for a different attribute of tours, while Table \ref{table:comparison} shows how the tours recommended by the algorithms compare according to the attributes.

\begin{table*}
  \caption{Tours comparison according to several attributes (rows one to four) and overall quality (rows five and six). Columns two to five indicate how many interviewees say that their \greedy{} tour features the attribute more than their \optimal{} tour (\greedy{} $>$ \optimal{}), less than \optimal{} (\greedy{} $<$ \optimal{}), approximately the same as \optimal{} (\greedy{} $\simeq$ \optimal{}), and the number of missing answers (N/A). The total number of interviewees is 16.}
  \label{table:comparison}
  \begin{tabular}{ccccc}
    \toprule
     & \greedy{} $>$ \optimal{} & \greedy{} $<$ \optimal{} & \greedy{} $\simeq$ \optimal{} & N/A \\
    \midrule
    Segue top-bottom bias & 7 & 0 & 0 & 9 \\ 
    Segue diversity & 9 & 1 & 4 & 2 \\
    Segue narrativity & 0 & 6 & 4 & 6 \\
    Song arrangement & 5 & 6 & 1 & 4 \\
    \midrule
    Segue interestingness & 9 & 5 & 2 & 0 \\
    Tour quality & 9 & 5 & 1 & 1 \\
  \bottomrule
\end{tabular}
\end{table*}

\subsubsection{Segue top-bottom bias} \label{theme:1.1}
Many interviewees mention that the first segue in a tour is more interesting than the last segue, which is a phenomenon we refer to as top-bottom bias. Among the 16 interviewees, 14 find that \greedy{} exhibits top-bottom bias, as we might expect. In \optimal{}, the trend is less clear, as we might also expect: seven interviewees find top-bottom bias, seven do not, and two interviewees cannot decide.
In total, seven interviewees say that the top-bottom bias is stronger in \greedy{} than in \optimal{} and the rest do not offer an opinion. The results corroborate the offline results in \cite{gabbolini2021play}, where \greedy{} is found to have top-bottom bias, and \optimal{} is not.

\cite{gabbolini2021play} also investigates how the interestingness of segues varies throughout the tour, referring to this as the \textit{outline}, finding that in \greedy{} the outline is falling and in \optimal{} it is not falling.
In our study here, the outline of \greedy{} is judged to be falling by six interviewees, bumpy by six, pyramidical by two, and flat by two. The outline of \optimal{} is judged rising by five interviewees, falling by four, flat by three, bumpy by two, pyramidical by one, and in one case we do not have an answer. So, the results corroborate only partially those in \cite{gabbolini2021play}, probably because the $interestingness$ function used in \cite{gabbolini2021play} to infer the outlines, and in the work at hand, only partially corresponds with what humans judge to be interesting, as we detail in Section \ref{theme:2.1}.

\subsubsection{Segue diversity} \label{theme:1.2}
The majority of the interviewees (14 out of 16) mention that the diversity of the segues in the tours, or \textit{segue diversity} for short, is an important attribute to have. (Notice that this attribute concerns the diversity of the segues, not of the songs.) Many interviewees mention that tours with low segue diversity would not be nice to listen to. For example, \p{8} says: \quot{then (the segues) start repeating, and so it gets less interesting as we go down}; and \p{16} says: \quot{I think if I was getting the same information \dots{} that would be becoming boring after a while, I \dots{} want diversity in information}. Many other interviewees make a similar point, that tours with higher segue diversity are nice to listen to. For example, \p{5} says: \quot{I just preferred the diversity, and it was like you didn’t know what was coming next, and what bit of information you were going to learn was. That was kind of nice, that you didn’t know!}; and \p{1} says: \quot{I feel like the information is \dots{} diverse \dots{} I’d say it keeps your attention more that way}. Most of the above comments refer to the diversity of segues that are relatively close to each other in a tour. That is, it is important for neighbouring segues to be diverse, while we have less guidance from our interviewees for segues that are more remote from each other. For example, \p{5} says: \quot{I felt like (the segues) grabbed my attention a bit more (in \greedy{}), because (in \optimal{}), the last 6 six were about genres … Maybe those six were also on (\greedy{}), but actually I did not notice as much}. 

Respectively three and four interviewees say that \greedy{} and \optimal{} have low segue diversity, while respectively six and five of them say that \greedy{} and \optimal{} have high segue diversity. We do not have an answer in the rest of the cases. The diversity is sometimes low because neither algorithm implements an explicit segue diversity mechanism.

Nine interviewees say that their \greedy{} tour is better than their \optimal{} in diversity, one says the opposite, four say they are not sure, and the rest do not offer an opinion.


The result may be explained by previous work, as \cite{gabbolini2021play} finds that the standard deviation of the $interestingness$ scores of the segues in \greedy{}'s tours is higher than in \optimal{}'s tours. That is, \greedy{} generates segues with a broader range of $interestingness$ values, some segues are very interesting and some are not, and segues with different interestingness values typically have different topics.

\subsubsection{Segue narrativity} \label{theme:1.3}
The narrativity of the segues in tours, or \textit{narrativity} for short, refers to the quality that sequences of segues in tours present a narrative with a coherent text. Very few interviewees (three out of 16) are in favour of narrativity, as two points against narrativity are made. Point (1) is that narrativity can correlate with low diversity, and low diversity is something to avoid, as mentioned in Section \ref{theme:1.2}. For example, \p{12} says: \quot{I don’t think (narrativity) really matters per se, like, if you had all (segues) linked to each other, it could get quite repetitive}. Point (2) is that the textual flow would not be easy to follow in any case with songs in between. For example, \p{6} says: \quot{I don’t think \dots{} that any kind of coherent narrative it’s something I’d look for \dots{} I don’t think it is all that important really.}; and \p{8} says \quot{I think (the segues) should be independent! Because if there was a flow there I’d forget what it was saying before while listening to the song, so I’d lose the flow! So I guess independent makes much more sense}.

Only two and six interviewees say that respectively \greedy{} and \optimal{} have high segue narrativity, while respectively nine and six of them say that \greedy{} and \optimal{} have low segue narrativity. We do not have an answer in the rest of the cases. The narrativity is oftentimes low because neither algorithm implements an explicit segue narrativity mechanism.


Six interviewees say that their \optimal{} tour is better than their \greedy{} tour in narrativity, four say they are not sure, and the rest do not offer an opinion.


\subsubsection{Song arrangement} \label{theme:1.4}
Many interviewees (13) give their opinion on what makes a good song arrangement. The majority of them (11) mention some notion of similarity, with a variety of different terms, reflecting the fact that music similarity remains a concept with many definitions \cite{jones2007human}. For example, they say that subsequent songs should have the same ``tempo'', ``tone'', ''mood'', ``energy'', ``melody'' and ``artist voice''.

Interviewees do not agree on whether song arrangement is important in tours or not, as eight of them say it is important, while six say it is not. Interviewees \p{6} and \p{7} from the former group respectively say: \quot{Oh definitely! No matter what the context, if you were putting a bunch of songs together, you are going to want some elements of musical flow between them} and \quot{you cannot set the tone in one tempo or a mood and change it rapidly \dots{} there’s nothing worse}. Interviewees from the latter group say that segues make song arrangement not matter. For example, \p{2} says \quot{with the segues \dots{} the order (of songs) doesn’t really matter, cause there’s still a connection}; and \p{12} says: \quot{you are not listening to this song, and the next song begins: you listen to this song, and there’s a break where some text is read out, then you listen to the next one \dots{} so no, I don’t think the (song) order matters}. Some interviewees even appreciate when subsequent songs are diverse. (Here, we are referring to song diversity, and not segue diversity.). For example, \p{4} says: \quot{it was interesting seeing the connections \dots{} (between) different songs \dots{} (like that two very diverse bands) both started in 2006, I suppose that was really interesting}; and \p{2} says: \quot{(the tours) are both pretty interesting \dots{} because \dots{} I tried to choose music that was diverse, so it's interesting to see (the segues)}. It may be that segues between diverse songs are more unexpected, hence more interesting, as the interestingness guideline (3) of Section \ref{theme:2.1} indicates. However, the interviews do not provide enough material to confirm this speculation.

Respectively six and seven interviewees say that song arrangement is not good in \greedy{} and \optimal{}. In fact, of course, neither algorithm optimises for song arrangement, but for segue interestingness. However, recent work found segue interestingness to be related to similarity \cite{gabbolini2021interpretable}, which could explain why some interviewees say they are happy with the song arrangements (eight interviewees for \greedy{} and seven for \optimal{}).

Interviewees do not agree on which algorithm is better in song arrangement, as five of them say that in \greedy{} the song arrangement is better than \optimal{}, six say the opposite, one cannot decide, and four interviewees do not offer any opinion. The result is not surprising as neither algorithm directly considers song arrangement.

\subsection{Tour quality} \label{theme:2}
We ask interviewees to compare the quality of the tours recommended by \optimal{} and \greedy{} according to segue interestingness and the overall tour quality. The results are in the subsections below, and in Table \ref{table:comparison}.

\subsubsection{Segue interestingness} \label{theme:2.1}
Interviewees mostly agree that \greedy{} has more interesting segues than \optimal{}: nine interviewees say so, five say the opposite, and two cannot decide. The result is perhaps surprising, as \optimal{} is supposed to maximise overall segue interestingness. 
But \optimal{}'s model is only partially accurate for two reasons, as noted in Section \ref{theme:3.1}: (1) the $interestingness$ function itself is only partially accurate; (2) the interestingness of a list of segues is not a simple sum; other factors intervene, for example the segue diversity. In particular, we suspect that \greedy{} outperforms \optimal{} because it is perceived to be more diverse, and diversity positively impacts segue interestigness, as noted in Section \ref{theme:1.2}.

\subsubsection{Overall tour quality} \label{theme:2.2}
Interviewees mostly agree that \greedy{} produces overall better tours than \optimal{}: nine interviewees say so, five say the opposite, one cannot decide, and in one case we do not have an answer. Some interviewees mention the reason behind their choice: nine of them mention more interesting segues; also nine participants mention better song arrangement. Note that interviewees were free to give multiple reasons, and some of them mention both segue interestingness and the song arrangement. We infer that \greedy{} is preferred to \optimal{} because of the segues, which are more interesting in \greedy{}, as noted in Section \ref{theme:2.1}, while song arrangement is equally good in both algorithms, as noted in Section \ref{theme:1.4}. The result also highlights the importance of song arrangement, which is mentioned by interviewees as a key factor when assessing tours. 

\subsection{Concept evaluation} \label{theme:3}
Interviewees evaluate tours positively, as the general opinion is that the sequences of songs and segues are nice to listen to. For example, \p{5} says: \quot{I’d be happy to listen to that (tour) \dots{} you learn something new about all tunes you like to listen to}; \p{7} says: \quot{I really like it! \dots{} it is nice to see how all songs are linked together, it is really cool!}; and \p{1} says: \quot{I like how the songs \dots{} were linked by pieces of information, something that they shared}. 
Some interviewees would like to see tours available in streaming services. For example, \p{8} says: \quot{I’d certainly use this prototype if it was implemented in a music streaming platform.}; \p{6} says: \quot{(the segues) would be quite interesting for a listener, if they were available in a streaming service}; and \p{5} says: \quot{I would like to see (tours) available, like they aren’t very available. It’d be nice to have the option (on streaming services) to switch them on and off. It is nice to listen to text, or to listen to something being spoken, in between listening to songs}.
Finally, the majority of interviewees (nine out of 16) mention that they would welcome a tour like those they were presented with.

The majority of interviewees find the segues in tours to be interesting, along a number of different dimensions. For example, \p{11} and \p{9} find segues to be non-trivial, as \p{11} says: \quot{It was a pretty cool (segue) actually. Like, I wouldn’t have thought about making a connection}; and \p{8} finds segues to be surprising \quot{being surprised, by interesting facts, like what this segues are producing here, that would make the listening experience more exciting}. More generally, \p{1-5}, \p{7}, \p{8}, \p{10-12} and \p{15} like the segues. For example, \p{15} says: \quot{I liked the connections, from top to bottom, like something that ties up everything}.
Some interviewees recognise the value of having segues. For example, \p{5} finds segues to be educational: \quot{I also liked the bits of information \dots{} It was kind of educational as well as listening to songs}; and \p{10} finds that segues add to the listening experience: \quot{(segues) add to the music, like you’d enjoy more your listening experience}. The above results provide some evidence that the $interestingness$ function, even though only partially accurate, is nevertheless related to user-centered perceptions of segue interestingess.

However, four interviewees report that segues are occasionally uninteresting. For example, \p{4} says: \quot{Some of the connections are a bit weaker \dots{} But I think it's mostly very interesting yeah!}; and \p{12} says: \quot{some of the links are quite tenuous, like two artists are both from the USA, that one felt a bit grasping}. These opinions agree with previous work, as \cite{gabbolini2021play} finds that the interestingness of segues in tours is not constant, but subject to considerable deviation. Hence, from \cite{gabbolini2021play} we expect some segues to be considerably less interesting than others, which is exactly what we report here.

Interviewees mention that tour quality is context-dependent, a similar point being made in \cite{Behrooz19}: tours are suited for active listening, \ie{} when attending to the music, and not passive listening, \eg{} when the music plays in the background to some other activity.
For example, \p{4} says: \quot{sometimes \dots{} you may just want to listen to your music in the background. But sometimes, if you want to sit down and think about the music your are listening to, (the segues) might be like an interesting, like a different, fun, new way to consume your music.}; and \p{9} says: \quot{a lot of times you listen to music as a background, whereas I think (for a tour you need) designated hours in your day, like I’m going to sit down and listen to this}.

\subsection{Tour improvements} \label{theme:4}
Interviewees identify a number of potential improvements for tours.

\subsubsection{Interestingness modelling} \label{theme:3.1}
Many interviewees say what interestingness means for them, and, from this, we extract four guidelines on how to distinguish interesting from uninteresting segues.

Guideline (1) is to be specific, as some interviewees say that specific segues are more interesting than general segues. For example, \p{1} says: \quot{I wouldn’t like to see information that seems kind of generic maybe \dots{} it just seems no care and attention has been put in creating that}. \p{7} gives an example of very general segue: \quot{The bit the says "one is based in the US and the other is based in the same country", because I think this is a way of linking nearly every one even, I like knowing a little bit more of information about someone}. The very same kind of segue is criticised also by \p{12}. Interviewees have a different opinion about specific segues. Again \p{12} says, \quot{the segues that are specific, are the most interesting, but the other like \dots{} these two people are from the same large country, it is a bit ... uninteresting}, and provides an example of a more interesting segue, which involve a more specific place: \quot{this group founded in (New) Jersey and also this other}.

Guideline (2) is to limit text length, as some interviewees say that long segues are typically undesirable. For example, \p{2} says: \quot{I would not want to see overly long segues}, and \p{14} says \quot{Not too much information, not too many words, otherwise I’d be like: please!}. We do not investigate how much text is too much for an interviewee. However, we report that the longest segue that was shown to \p{14} was of 22 words, while for \p{2} it was of 19 words.

Guideline (3) is to be aware of prior knowledge, as some interviewees say that segues that contradict their prior knowledge are to be avoided. For example, \p{15} says \quot{apart from that (same genre segue) on Christmas music, which is not correct. Apart from that, I enjoyed it}. Also, many interviewees say that known segues are uninteresting. For example, \p{14} says \quot{One of the segues was U2 and Fionn Reagan are Irish and \dots{} I know that because I’m Irish! \dots{} that is just tedious and boring.}; and \p{12} says \quot{the rest (of the segues) was kind of … I knew them already \dots{} so I’d be kind of "ehm, okay"}. Finally, many interviewees say that novel/unexpected segues are interesting. For example, \p{12} says: \quot{I didn't know some of this stuff, such as (these two bands) being from the same state, it’s interesting}; and \p{16} says: \quot{(in good segues) you’d have kind of information that you didn’t know and you were just learning. Something interesting, like one of those of moments like oh my god I didn’t know that these two bands were connected}.

Guideline (4) is to avoid controversies, as some interviewees point out that controversial segues might be typically undesirable, such as those involving artists personal lives. For example, \p{11} says: \quot{(I don't want to hear) scandals related to the artists \dots{} It’d ruin the listening experience}. Similarly, \p{8} says \quot{(I wouldn't want to hear something) too particular like, when this artist was married, or when this artist was in prison}. 

Only two of the guidelines above match the $interestingness$ function used by the algorithms in this work. This function, defined in \cite{gabbolini2021generating}, combines rarity, which arguably entails specificity, and shortness. 
Therefore, \greedy{} and \optimal{} strive to maximise an imperfect $interestingness$ function, that could be improved by considering all four of the guidelines above. Even so, $interestingness$ does a good job in distinguishing actually interesting segues from others, as reported in Section \ref{theme:2.1}. Moreover, in this work we model the interestingness of a list of segues as the sum of the $interestingness$ function applied to the individual segues in the list. This will only approximate human perceptions of interestingness because also other factors matter. For example, diversity of neighbouring segues makes the tour more interesting overall. One improvement is to correct the estimate of the interestingness of a list of segues by taking into account also diversity, and especially the diversity of nearby segues in tours, as noted in Section \ref{theme:1.2}.

\subsubsection{More biographical segues, less genre segues.} \label{theme:3.2}
Many interviewees highlight examples of interesting and uninteresting segues seen in the tours. One class of interesting segues are (factual) biographical segues about the artists. For example, \p{5} likes segues about awards won by artists; \p{9} likes segues about locations, such as those about where an artist was born; \p{3} likes segues about live events, such as in which festival an artist performed; and \p{9} likes segues about dates, such as when an artist was born. Interviewees would like to see even more biographical segues, including ones that are not available in tours at the moment. For example, \p{12} says they would like segues to draw from music news databases, that gather information such as emerging artists listings, recording studios, and albums release dates.
One class of uninteresting segues are (musical) genre segues, such as the ``jazz'' segue in Figure \ref{fig:toy_sample} (c). For example, \p{1}, \p{3}, \p{4}, \p{6}, \p{12}, \p{15} say that genre segues are not likeable and uninteresting.

We believe that biographical segues and genre segues are perceived as respectively interesting and uninteresting at least in part because of interestingness guideline (3) of Section \ref{theme:2.2}: to be aware of prior knowledge. Biographical segues, as facts, match guideline (3): they are not likely to contradict user prior knowledge, unless the user is wrongly informed, and specific facts are likely to be novel/unexpected for non-expert users. For example, the segue ``Flying Lotus is the grand-nephew of Alice Coltrane'' of Figure \ref{fig:toy_sample} (b) cannot contradict user prior knowledge, unless the user is wrongly informed about the relation between the two artists, and they are likely to be novel/unexpected for the user, unless the user is an expert in artists' biographies. 
Genre segues, instead, may often not match guideline (3): they can contradict user prior knowledge, as people are found to often disagree on music genres \cite{sordo2008quest}, and they can be known to the user, as genres are one of the most common ways in which people characterise music \cite{petridis2022tastepaths}. Interviewees agree with our interpretation, as \p{16} is disappointed by a genre segue contradicting their prior knowledge: \quot{I’d like not to grow a question that would put me off \dots{} you might thinking about that instead of enjoying the music \dots{} that Leonard Cohen (segue) is a good example \dots{} I started thinking: is he rock music?}; while \p{9} says they know genre segues already: \quot{There were someones that were linked by same genre \dots{} and maybe that’s something you could have come up with yourself \dots{} I suppose that is something you have an idea of already}; \p{8} also thinks that genre segues are very trivial: \quot{those (genre) segues have problems, they are \dots{} very trivial}.

Biographical and genre segues belong to the broader groups of respectively factual and opinionable segues, which we suspect to be more and less interesting, again because of interestingness guideline (3): factual segues are not likely to contradict user prior knowledge, unless the user is wrongly informed, while opinionable segues, by definition, are more likely to do so. However, future work is needed to verify our supposition. Some factual segues that could be included in tours are suggested by the interviewees. For example \p{14} would like segues drawn from artists' biographies, both related and not related to music, \eg{} which and how many pets an artist has; \p{7} would like acoustical information segues, such as the tempo or key of a song; and \p{9} would like lyrics segues, such as which keywords features in the song lyrics.

Summing up, segues in tours could be improved by including more biographical information and by avoiding musical genres. Another potential improvement is to include more factual information, such as those in the last paragraph, and avoid opinionable statements. The topics of segues can be modified by altering the knowledge graph available to algorithms, and described in Section \ref{sec:method}, to include the information that should feature in segues, and to exclude the information that should not feature in the segues.

\subsubsection{Song selection} \label{theme:3.3}
The algorithms we are using are given a set of songs as input, and return a sequence made from this same set of songs, broken up by segues. That is, the algorithms are not designed to perform any song selection. In this work, we ask interviewees to send us ten songs they are familiar with, that we input to the algorithms, as described in Section \ref{sec:experiment}.
In this Section, we report what interviewees say about how to appropriately select the songs in input to the algorithms. In summary, we find that the right level of familiarity is key.


The tours we show during the interviews feature familiar music only. Many interviewees mention that they would like tours to feature some unfamiliar songs, along with familiar songs, a concept we refer to as UF-tours (Unfamiliar/Familiar tours). A UF-tour is similar to popular features of music steaming services, such as Spotify's \textit{daily mix}\footnote{\url{https://newsroom.spotify.com/2018-05-18/how-your-daily-mix-just-gets-you/}}, which is a playlist of familiar and unfamiliar music. A UF-tour, however, is different from a \textit{daily mix} because the unfamiliar songs are introduced by a segue. The majority of interviewees (10 out of 16) say they would welcome a UF-tour. For example, \p{12} says \quot{"Spotify would give your daily mix (that is) songs you know mixed with similar songs \dots{} I would have never listened to. (A UF-tour) can be quite interesting actually, because it creates a connection, it is not just a vague music floating around \dots{} having that little bit of information to introduce the artist and the actual songs would be quite interesting I think, I would be you more inclined to go to the actual artist page and listen to more of their music."}; and \p{3} says: \quot{I'd be very interested in this, because on Apple Music, if you set auto-play from your songs, they will bring on just a random song \dots{} while this will give you why they are similar, what connection do they have.}
It is not clear whether UF-tours are preferred over F-tours (tours of familiar music only), because not enough interviewees explicitly compared the two concepts. Interviewee \p{5} suggests a button to switch between an F-tour and a UF-tour, which can be a sound design choice before future investigation on the subject.

Many interviewees comment on tours that feature only unfamiliar music, a concept we refer to as U-tours. The general opinion is mostly negative, as no interviewee says that they would welcome a U-tour. The main reason is that segues between two unfamiliar songs lose their meaning. For example, \p{9} says \quot{(segues are valuable) with music that is important to you. If that music is not important to you in the first place, I think (the segue) it's just going to be semantic information \dots{} it wouldn’t bring any value to me}. Familiar songs are needed to keep interest in the tour alive. For example, \p{12} says \quot{I think the mix of familiar and novel songs works, because the familiar songs are kind of an anchor to you, whereas with two unfamiliar songs \dots{} (a segue) wouldn’t mean much to me}. The above result also suggests avoiding segues between two unfamiliar songs in UF-tours.

One difficulty in implementing UF-tours and U-tours is recommending which unfamiliar music to display in the tour. Interviewees suggest that recommended music should fit the user's tastes. For example, \p{2} says \quot{having music you don't like \dots{} is kind of inevitable when finding new music but ideally a tour would have music you will like in advance \dots{} having a tour that doesn’t read your preference very well \dots{} wouldn’t be good}. Moreover, it may be important that all songs in a tour are similar to each other, as noted in Section \ref{theme:1.4}.

\subsubsection{Presentation} \label{theme:3.4}
Tours are presented to interviewees through two mediums, visual and aural, as explained in Section \ref{sec:experiment}. The visual presentation is similar to Figure \ref{fig:toy_sample} (b) and (c), while the aural presentation is a sequence of songs, alternated by segues, read by a TTS engine. It is not clear which medium is preferred, as we do not investigate the matter explicitly. 

Interviewees do however identify a number of flaws in the presentation of the tours.
For example, the TTS is perceived as too ``robotic'', not human-like, and it is not liked by some interviewees. For example, \p{6} says: \quot{(would have been better) if the audio links were not spoken as obviously by a computer, you know if it was more human-like, more engaging}.
Also, some interviewees do not like the short pauses between the music and the segues, and would like the song and the segue to overlap a little, similar to radio programs. For example, \p{5} says: \quot{you could merge the songs a little bit (so that) the voice over came in towards the end of the first song, while the second song began while the voice over is running, eliminating that downtime between the two}.
Finally, some interviewees question the rule of having a segue between every pair of songs, and suggest that, if not interesting enough, a segue could be skipped. For example, \p{16} says: \quot{If the information isn’t going to be something that’d grab, I wouldn’t want a segment, you-know? If you can’t find an interesting enough segment, you will just have to not have a segment}.

\section{Conclusion}
In this Section, we summarise the interviews reported in Section \ref{sec:results} to answer all the \q{}s we pose, we discuss the limitations of the work that we have presented here, and we outline possible future work on tour recommendation.

\subsection{Discussion of \q{}s} \label{subsec:discussion}
\textbf{\q{1}. What are the attributes of tours?}
(1) According to the interviewees, tours recommended by \greedy{} feature a segue top-bottom bias, that is the first segue tends to be more interesting than the last segue, while tours recommended by \optimal{} do not. (2) Tours do not always feature segue diversity, but tours recommended by \greedy{} are found to be more diverse than tours recommended by \optimal{}. (3) Tours do not always feature segue narrativity, that is the quality that the sequence of segues in tours presents a narrative with a coherent text, but tours recommended by \optimal{} are found to feature more narrativity than tours recommended by \greedy{}. (4) Tours do not always feature good song arrangement, and tours recommended by \greedy{} are found to be equally good in song arrangement to tours recommended by \optimal{}.

\textbf{\q{2}. What attributes of the tours are good/bad?}
Segue diversity is definitely a good tour attribute, as 14 out of 16 interviewees say so. Song arrangement is another good attribute, as interviewees say that good song arrangement is a reason as important as good segues to prefer a tour over another. Segue narrativity, instead, is considered a good attribute of tours by only three interviewees. In conclusion, segue diversity and song arrangement appear to be the two most important attributes of the tours investigated in this work. 

\textbf{\q{3}. Which algorithm recommends better tours, and why?} 
\greedy{} seems to recommend higher quality tours than \optimal{}. Interviewees motivate their choice mentioning higher segue interestingness and better song arrangement. We infer that \greedy{} is preferred to \optimal{} because of segue interestingness, which is higher in \greedy{}, while the quality of song arrangement is similar in both algorithms. In turn, segues in \greedy{} could be more interesting because of the higher diversity, which is regarded as one important characteristics of segues. 

\textbf{\q{4}. How valuable is the concept of tours in general?} 
The majority of interviewees find that the algorithms recommend tours with interesting segues, and say they would welcome a tour such as those they were presented with. However, accepting tours is dependent on context: tours require active listening, and they are not suited for passive listening.  In conclusion, participants mostly evaluate the concept of tours positively in general.

\textbf{\q{5}. What are possible improvements to tours?}
(1) Take into account user prior knowledge and controversial content when scoring the segues. In particular, known segues, segues contradicting user prior knowledge and controversial segues should have low scores. (2) Include more segues with biographical information and fewer segues about musical genres. (3) Choose music carefully, admitting, from time to time, unfamiliar music, always chosen to fit the user's tastes. 
(4) Mind the presentation, employing an appropriate TTS engine to read the segues, avoiding silences by overlapping songs with segues, and skipping segues if not interesting enough. (5) Implement a good song arrangement mechanism based on song similarity, as song arrangement is a fundamental attribute of tours, but it is not considered by the algorithms we use. (6) Implement a segue diversity mechanism, as segue diversity is a fundamental attribute of tours, but it is not considered by the algorithms we use.

\subsection{Limitations} \label{subsec:limitations}
Our work has limitations that should be acknowledged. We conduct semi-structured interviews, as they suit best our need of answering the \q{}s, as argued in Section \ref{sec:experiment}. However, this kind of experimental setting could affect the validity of some results. For example, when evaluating the concept of tours (\q{4}), interviewees could infer that tours are the work of the interviewers, and may feel pressure to provide a positive evaluation. We attempt to relieve the pressure, as explained in Section \ref{sec:interview_guide}. However, we cannot prove the effectiveness of the attempt.

Moreover, semi-structured interviews are a high cost protocol, which constrains the number of participants \cite{braun2006using}, 16 in our case. 
In our case, we resort to a convenience sampling of university students, which does not imply that our result extends to a broader sample of the population.
Moreover, the small number of participants limits the statistical power of the experiment, as we cannot afford to have more than two treatments \cite{knijnenburg2015evaluating}. In our case, we fix the two treatments to be the two tour recommendation algorithms, as we explain in Section \ref{sec:intro}. Therefore, we do not include any playlist recommendation algorithms among the treatments, so we cannot assess whether tours are preferable to playlists or not. By analysing the results of Section \ref{sec:results}, we may infer that playlists are preferable to tours when listening to music passively, and that tours can be preferable to playlists when listening to music actively, which corroborates previous work \cite{Behrooz19}. However, at this stage this is more an intuition than a definite answer. Similarly, we do not include other interesting treatments in the experiment, such as a tour recommendation algorithm which account for segue diversity, such as the one proposed in \cite{Behrooz19}. In any case, the work of \cite{Behrooz19} is not fully reproducible, as some fundamental parameters of their algorithm, such as the author-defined segue importance weights, are not shared in the paper.

Finally, while the interviews hint at several guidelines for the construction of functional tour recommender systems, we cannot forecast the impact of those guidelines, and their effectiveness can be only assessed by comparing a tour recommender that implements those guidelines and a tour recommender that does not.

\subsection{Future work} \label{subsec:future_work}
The answers to \q{5} that we present in Section \ref{subsec:discussion} are a source of ideas for future research on tour recommendation. One idea is to develop a mechanism to mine the user's prior knowledge, which could be done explicitly by prompting the user for opinions on segues, which is similar to active learning \cite{RubensRecSysHB2010}, or implicitly, by building a personalised classifier. Another idea is to develop a mechanism to detect controversial content in segues, which could take inspiration from work on controversial text detection \cite{aharoni2014benchmark}. The two above ideas are necessary to build a better model of segue interestingness, as highlighted in Section \ref{theme:3.1}. Another research direction is building a tour recommender able to surface the right music, as current tour recommenders do not have this feature: they assume they are given the right songs for the tours. Future work could also include an on-line experiment, such as an A/B test, so as to provide an evaluation of the concept of tours in a natural environment.

\begin{acks}
This publication has emanated from research conducted with the financial support of Science Foundation Ireland under Grant number 12/RC/2289-P2  which is co-funded under the European Regional Development Fund. For the purpose of Open Access, the author has applied a CC BY public copyright licence to any Author Accepted Manuscript version arising from this submission.
\end{acks}

\bibliographystyle{ACM-Reference-Format}
\bibliography{sample-base}

\end{document}